\documentclass[letterpaper, 10 pt, conference]{ieeeconf}  
\usepackage{graphicx}
\usepackage{multirow}
\usepackage{subfig}
\graphicspath{{figs/}}

\IEEEoverridecommandlockouts                              

\newcommand{\cross}[1]{#1 $\times$ #1}

\title{\LARGE \bf
Deep Feature Learning from a Hospital-Scale Chest X-ray Dataset with Application to TB Detection on a Small-Scale Dataset 
}

\author{Ophir Gozes$^{1}$ and Hayit Greenspan$^{2}$
\thanks{$^{1}$Ophir Gozes is with Faculty of Electrical Engineering,
       Tel Aviv University, Israel
        {\tt\small ophirgozes@mail.tau.ac.il}}%
\thanks{$^{2}$Hayit Greenspan is with the Department of Biomedical Engineering, Tel Aviv University
        {\tt\small  hayit@eng.tau.ac.il}}%
}

\begin{document}

\maketitle
\thispagestyle{empty}
\pagestyle{empty}

\begin{abstract}
The use of ImageNet pre-trained networks is becoming widespread in the medical imaging community. It enables training on small datasets, commonly available in medical imaging tasks.
The recent emergence of a large Chest X-ray dataset opened the possibility for learning features that are specific to the X-ray analysis task. In this work, we demonstrate that the features learned allow for better classification results for the problem of Tuberculosis detection and enable generalization to an unseen dataset. 

To accomplish the task of feature learning, we train a DenseNet-121 CNN on 112K images from the ChestXray14 dataset which includes labels of  14 common thoracic pathologies. In addition to the pathology labels, we incorporate metadata which is available in the dataset:  Patient Positioning, Gender and Patient Age. We term this architecture MetaChexNet. As a by-product of the feature learning, we demonstrate state of the art performance on the task of patient Age \& Gender estimation using CNN's. Finally, we show the features learned using ChestXray14 allow for better transfer learning on small-scale datasets for Tuberculosis.

\end{abstract}

\section{Introduction}
The recent emergence of large x-ray datasets has opened the way for the development of Computer-Aided Detection (CAD) tools for a set of the most common chest pathologies (ChexNet \cite{c4}, Wang \cite{c6}). For the case of other pathologies, such as for Tuberculosis (TB), small datasets still remain a challenge.
 According to the World Health Organization Global Tuberculosis report 2018 \cite{c13}, TB is one of the top 10 causes of death worldwide.  
 In 2017, TB caused an estimated 1.6 million deaths worldwide.
 If detected in early stages TB can be treated, thus there is a need for the development of CAD tools for automatic screening of TB \cite{c13}.
 
Previous work by Lakhani et al. \cite{c11} demonstrated the advantage of using ImageNet \cite{c10} pre-trained architectures for TB detection on small-scale datasets. The strategy of using the ImagneNet pre-trained network is effective, since lower level natural-image features can be relevant to medical images. This was further verified by Sivaramakrishnan et al. \cite{c12}. In the current work,  we show that although clearly important, this transfer is sub-optimal.
In addition to ImageNet based features, we propose to further fine-tune an ImageNet pre-trained network with the hospital-scale ChestXray14 dataset. This further adapts the features learned to work on medical chest X-ray images. 


 ImageNet pre-trained networks are trained for classification of  1.2 Million natural color images into 1000 classes as part of the ILSVRC challenge \cite{c14}. To train our feature extraction network, which we term MetaChexNet, we start with an ImageNet pre-trained architecture and further train it on the ChestXray14 dataset that contains 112K images with the associated labels of 14 common thoracic pathology labels; In addition to the 14 given labels, we include the metadata which is comprised of patient age, position, and gender. 

\vspace{0.1in}
The contribution of this work includes the following:
\begin{itemize}
	
	\item We present a feature learning scheme which uses pathology labels and metadata of a hospital-scale chest X-ray dataset.
	
	\item  We demonstrate a method for gender, age and position estimation on Chest X-ray using deep learning.

	\item We demonstrate the applicability of features learned from a hospital-scale dataset to tackle small-scale chest X-ray dataset for the case of Tuberculosis. We compare our results to ImageNet pre-trained architecture.
	
\end{itemize}
The proposed method is presented in Section 2. Experimental results are shown in Section 3. In Section 4, a discussion of the results is presented.

\section{Methods}
\begin{figure}[h]
	\centering
	\fbox{\includegraphics[width=2.8in,trim={1.4cm 5cm 2.4cm 1cm},clip]{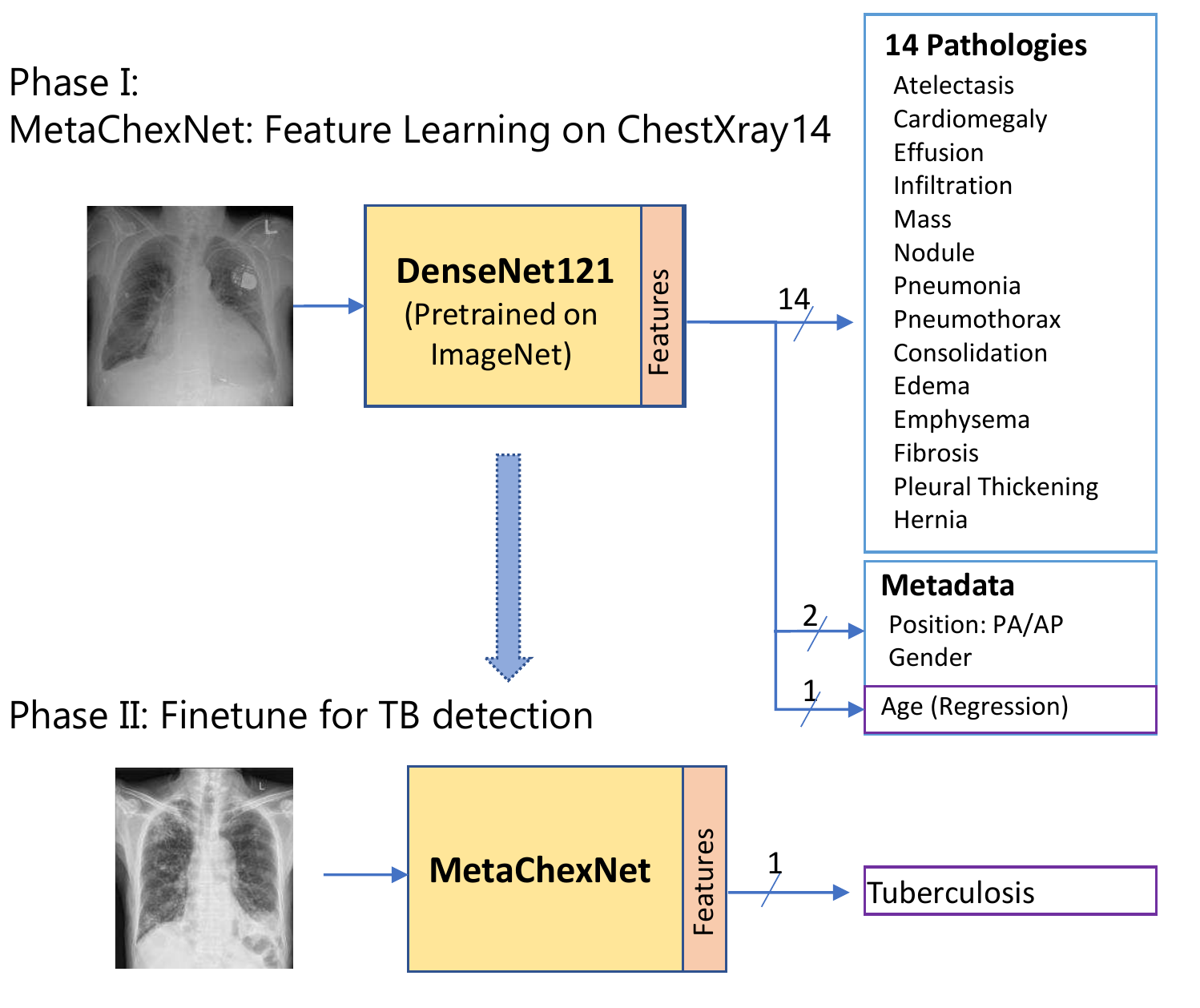}}
	\caption{MetaChexNet Training Phases}
	\label{metaxchexnet_training}
\end{figure}

 Our method is comprised of two phases, as shown in Figure  \ref{metaxchexnet_training}. 
In Phase I, we learn Chest X-ray specific features. In Phase II we fine tune the network trained in Phase I (MetaChexNet) for TB detection. 

	\begin{table*}[!hbtp]
		\centering
			\begin{tabular}{c|c|c|c|c}
				\hline
				Layer Name                                                                          & Size     & Connected to & \multicolumn{1}{c|}{MetaChexNet}                                                                     & \multicolumn{1}{c}{ChexNet \cite{c4}} 
					\\ \hline
				Features   & 1024  &Last Dense Block output &\multicolumn{2}{c}{\cross{7} global average pool }                                                                                                                                                                 
				\\ \hline
				14 Pathologies Classification           & 14 &Features  & \multicolumn{2}{c}{14D fully-connected, sigmoid }                                                                                                                                                                                                                                                                                                                                                                                                    \\ \hline
				Position \& Gender Classification    & 2 & Features & 2D fully-connected, Sigmoid & \     -                                                                               \\ \hline
				Intermediate (internal)    & 10 & Features & 10D fully-connected, ReLU & \     -                                                                                                                                                                \\ \hline
				Age Regression     & 1 & Intermediate & 1D fully-connected, Sigmoid & \     -                                                                                                                                                                                                        \\ \hline
				\end{tabular}

		\vspace{1 ex}
		\caption{DenseNet121 based architectures for ChestXray14.    }
		\label{Metachexnet_structure}
		\vspace{-3 ex}
	\end{table*}

\subsection{Phase I: Chest X-ray Feature learning}

 In order to learn discriminative CXR specific features, we start with an ImageNet pre-trained DenseNet121 \cite{c5} architecture and train it on the NIH ChestXray14 dataset. We train the network for the multi-task problem of predicting pathology labels and the metadata associated with each image, as follows:
	\textbf{Task 1: Learning Chest X-ray pathology related features}-
The network is trained to perform binary multi-label classification of the 14 most common chest pathologies: \{Atelectasis; Cardiomegaly; Effusion; Infiltration; Mass; Nodule; Pneumonia; Pneumothorax; Consolidation; Edema; Emphysema, Fibrosis; Pleural-Thickening; Hernia\}.
	\textbf{Task 2: Learning metadata related features}-
The network is trained for metadata prediction which includes binary classification of the patient's position \{AP-Anteroposterior, PA-Posterioranterior\}, gender \{Female, Male\} in addition to age regression\{0-100\}.
 
\textbf{Architecture \& Training: }
Rajpurkar et al first used  DenseNet121 architecture on the ChestXray14 to perform multi-label pathology classification and termed it ChexNet \cite{c4}.
In our work we focus on DenseNet121 architecture with compression factor $\theta=0.5$. As the feature vector, we regard the output of the last average pooling layer of the DenseNet121 which is of size 1024. In comparison to ChexNet, we increase the size of the binary output vector to 16 to accommodate for the two extra binary metadata labels (position, gender). In order to allow for age regression, we add an intermediate dense layer followed by a sigmoid activated neuron.
We term the trained network MetaChexNet since it is an extension of the ChexNet architecture providing metadata prediction. The Network architecture is specified in Table  \ref{Metachexnet_structure}.
For the loss function, we use binary cross entropy loss for the binary variables and mean absolute error loss for the continuous age variable.
Training is performed with a Nesterov Adam optimizer (Nadam) with batch\_size=32 and an initial learning rate of 1E-3.
Learning rate is reduced by a factor of 10 each time validation loss stops improving after one epoch. As pre-processing, the images are re-sized to size $224\times224$  and converted to a three channel RGB image by channel duplication. The images are normalized by subtracting the ImageNet mean and dividing by ImageNet standard deviation. The age metadata was scaled to the range of [0,1].
For data augmentation, we randomly flip the training images horizontally 50\% of the time.

\subsection{Phase II: Fine Tuning for TB detection}

We address the task of TB detection using the features learned in Phase I. The dataset available for TB training is two orders of magnitude smaller then the NIH ChestXray dataset, thus it is more difficult to train deep architectures without pre-training.
To tackle the smaller size of the dataset, we employ data augmentation which includes horizontal flipping, random scaling $(scale factor = [0.9,1.1])$, and random multiplication by a constant$(range = [0.8,1.3])$. Each augmentation method is performed with a probability of 0.5.

Taking advantage of the features learned during phase I of the training, our TB detection network is constructed on top of MetaChexnet or ChexNet feature layer. We use a single sigmoid activated neuron for the TB class. 
During the training for TB detection, we finetune the entire network with the same hyper-parameters as used in Phase I. The loss function is binary cross entropy loss. We select the model that received the best AUC value on the validation set and use it for calculating the test metrics on the test set. 

   \begin{figure*}[!hbtp]
      \centering
      \framebox{\parbox{5.2in}{

      \includegraphics[width=5.2in]{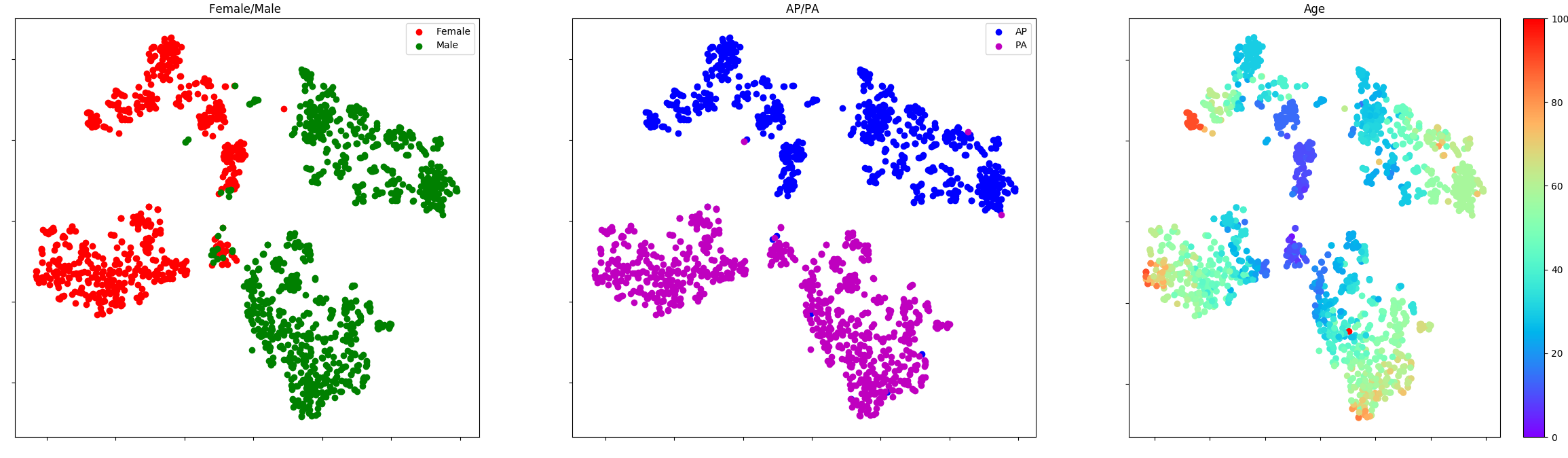}}}
      \caption{t-SNE: MetaChexNet feature visualization for Gender, Position and Age}
      \label{tsne_fig}
      	\vspace{-3 ex}
   \end{figure*}
   \section{Experiments and results}
\subsection{Chest X-ray Feature Learning}
For this experiment, the ChestXray14 dataset which contains 112K Chest X-ray images was split to three sets Train(104,266), Validation(6,336) and Test(1,518) with no patient overlap between the sets. The number of patients per set was 28,744, 1672 and 389, respectively.
The 14 pathology labels present in the dataset were text-mined from radiology reports. The information relating to age, gender and position was extracted from the image metadata.

Summary of the results of ChexNet and MetaChexNet is given in Table \ref{auc_nih}. We specify the results of both validation and test set. 
The Mean AUC for the detection of 14 pathologies is similar in both architectures. The AUC for gender and position was remarkably high ($>$0.99). MetaChexnet was able to predict patient age with an absolute error of 4.3 years.

\begin{table}[htbp]
\caption{Phase I Results on ChestXray14 (Validation,Test) }
    	\vspace{-3 ex}
\label{auc_nih}
\begin{center}
\begin{tabular}{l|c|c}
\hline
  & MetaChexnet &ChexNet \cite{c4}\\ 
\hline\hline
AUC 14 pathologies   & 0.83,0.80 & 0.83,0.80 \\ \hline
AUC Gender   & 0.997,0.996 & - \\
AUC Position(PA/AP)   & 0.998,0.998 & - \\
AGE  Error  &$-0.05\pm5.6Y$, $-0.16\pm5.68Y$  & - \\
AGE Absolute Error & 4.28Y,4.24Y &  - \\

\hline
\end{tabular}
\end{center}
\end{table}

\begin{figure}[htpb]
     \centering
     \framebox{\parbox{2.7in}{
     \includegraphics[width=2.7in,trim={0.5cm 0.1cm 0.7cm 0.4cm},clip]{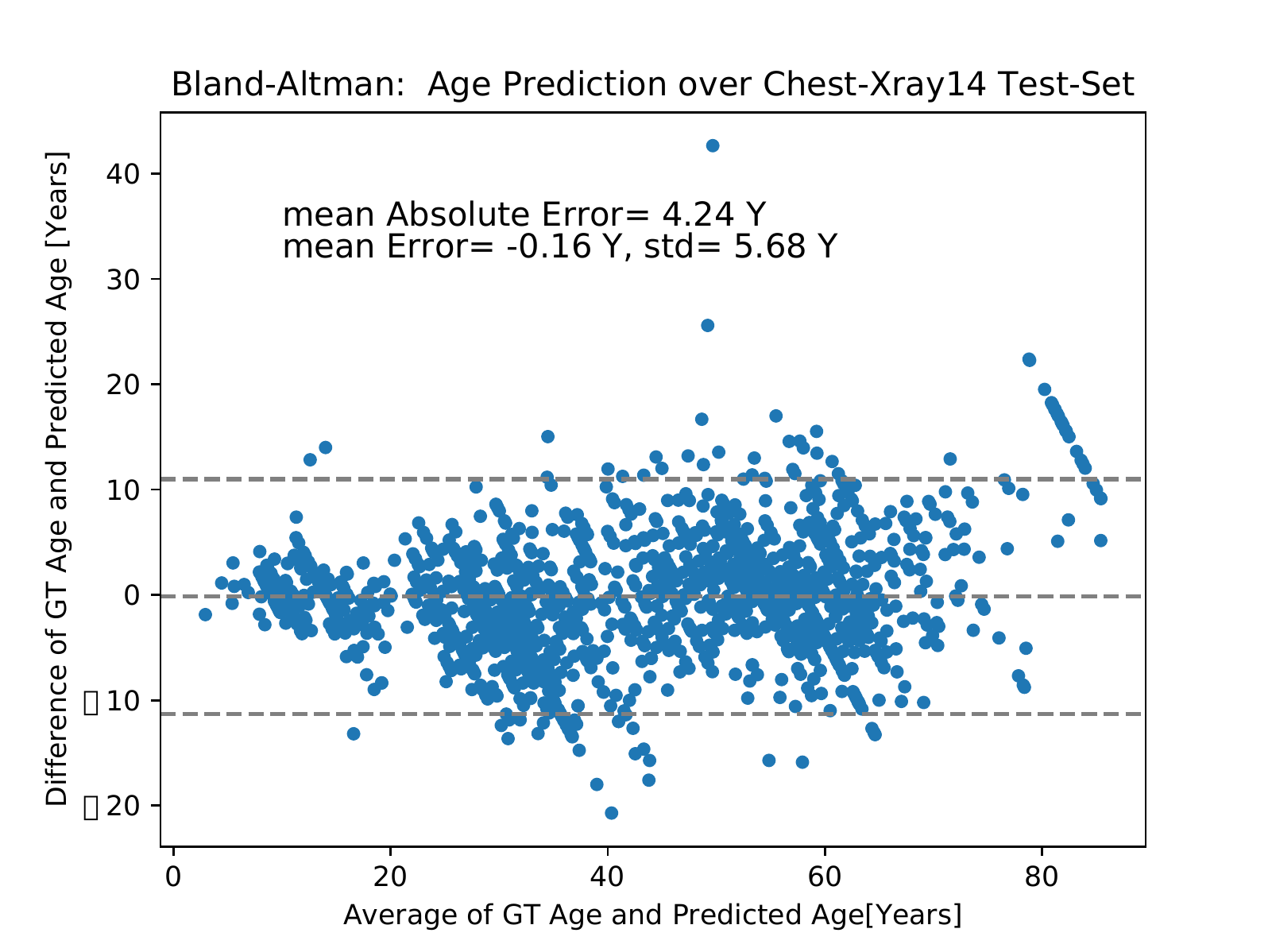}}}
     \caption{Bland Altman for Age distribution over NIH ChestXray14 test-set }
     \label{bland_altman}
  \end{figure}

In Fig.\ref{bland_altman} we display the Bland-Altman plot corresponding to age prediction on the ChestXray14 test subset. The age distribution among our test set was 42$\pm$18 years. It can be seen that the bias is small. 
  
\begin{figure} []

    \centering
  \subfloat[\label{1aa}]{
       \includegraphics[width=1.4in,trim={0.2cm 0.2cm 0.9cm 1.2cm},clip]{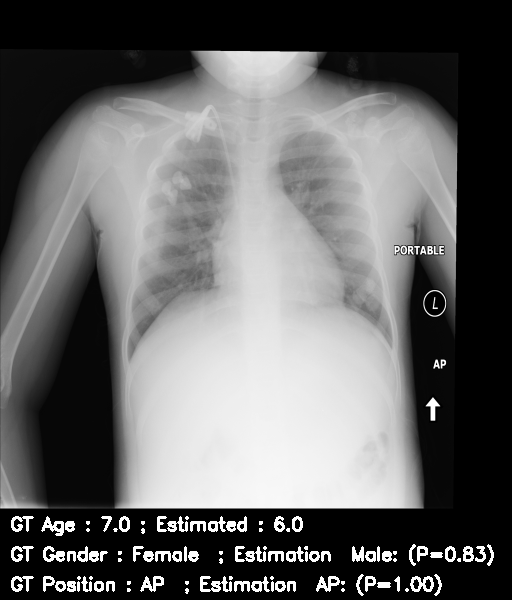}}
    \hfill
    \\
    \vspace{-2 ex}
  \subfloat[\label{1bb}]{
    \includegraphics[width=1.4in]{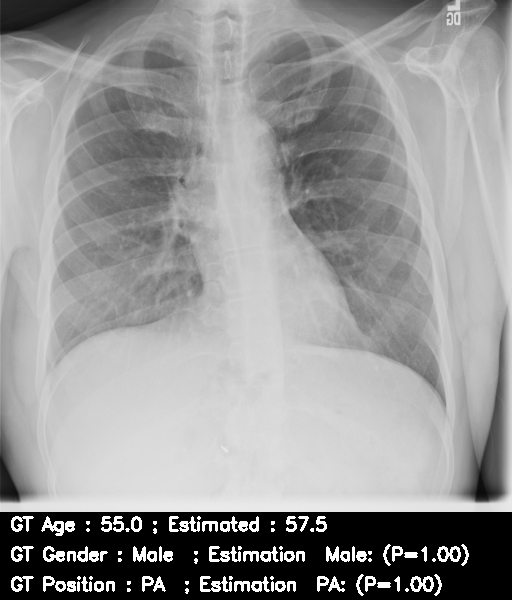}}
    \subfloat[\label{1cc}]{
    \includegraphics[width=1.4in]{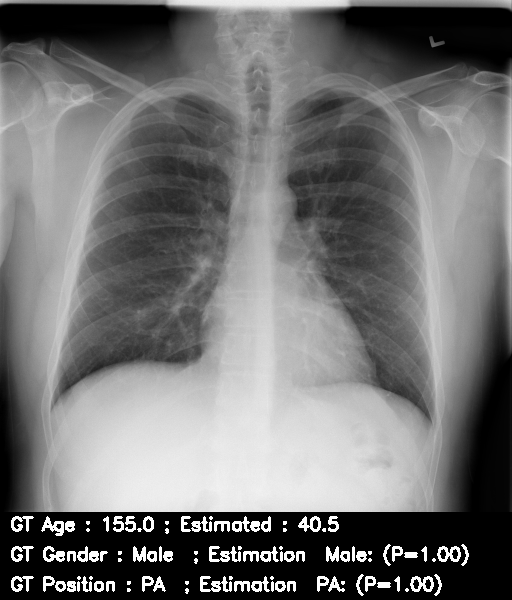}} 
  
  \caption{ MetaChexNet predictions on ChestXray14 \cite{c6}  test-set (P corresponds to the confidence of the prediction): (a) Gender Miss classification, (b) Correct classification (c) Wrong age label detection (patient \#27989) }
  
  \label{patient_photo} 
\end{figure}

{\it Feature Visualization}: 
To visualize the features learned by the MetaChexnet, we use t-SNE \cite{c9} visualization of the feature layer. In Fig.\ref{tsne_fig} we display the t-SNE corresponding to the feature representation of the test samples. It can be noted that the features are arranged in four clusters corresponding to gender and position. Viewing the Gender t-SNE, it is interesting to note that most of the gender classification mistakes occur in young ages (center of 4 clusters). An example of this type of miss-classification is visually demonstrated in Fig.\ref{patient_photo}-A.

A few example predictions are displayed in Fig.\ref{patient_photo}.
Using our algorithm we detected an erroneous age label which corresponds to patient ID 27989. The MetaChexNet age estimation was 40.5Y while the labeled age was 155Y (Fig. \ref{patient_photo}-C).

{\it Relationship between ChestXray14 pathologies and TB: }
Several of the pathology labels in the NIH ChestXray14 can appear as radiographic manifestations of TB \cite{c8}. To study the connection between ChestXray14 pathologies and TB, we used the trained ChexNet of phase I.  We predicted the 14 pathologies on Shenzen dataset and performed logistic regression on the log odds of the output probabilities. In Table \ref{log_res_coefficients} we show the Logistic Regression coefficients for each pathology. It can be noted that Fibrosis, Infiltration, Nodule, and Effusion were positively correlated with the presence of TB. The accuracy of the fit on the entire Shenzen dataset was 0.858. This motivates the use of ChestXray14 for discriminative feature learning.

\begin{table}[htbp]
\caption{Logistic regression coefficients for TB detection }
    	\vspace{-3 ex}
\label{log_res_coefficients}
\begin{center}
\begin{tabular}{l|c|l|c}
\hline
 Pathology & coefficient & Pathology & coefficient\\
\hline 
Atelectasis &  -0.76 &Pneumothorax &  -0.01, \\
Cardiomegaly &  -0.49 & Consolidation &  -0.011, \\
Effusion &  \textbf{0.86} &Edema &  0.21  \\
Infiltration &  \textbf{1.12} &Emphysema &  0.40\\
Mass &  0.056 & Fibrosis &  \textbf{1.24} \\
Nodule &  \textbf{0.83} &Pleural Thickening&  -0.67 \\
Pneumonia &  -0.72 &Hernia &  0.00\\

\hline
\end{tabular}
\end{center}
\end{table}

\subsection{Fine Tuning for TB detection }
In order to train for TB detection, the Montgomery and Shenzen datasets containing postero-anterior (PA) chest radiographs were used \cite{c7}. The two datasets contain normal and abnormal chest X-rays with manifestations of TB and include radiologist readings.
The Montgomery dataset was kept exclusively for testing, allowing examination of the generalization ability of the different networks. The images in Shenzen dataset used for training our algorithm were captured using Digital radiography machines (DR) while the images in Montgomery dataset were acquired using Computed radiography machines (CR). The composition of the datasets is given in Table \ref{datasets}.
 
\begin{table}[h]
\caption{TB Datasets}
\vspace{-3 ex}
\label{datasets}
\begin{center}
\begin{tabular}{l|c|c}
\hline  
  & TB Negative & TB Positive  \\
\hline
Shenzen-Training  & 226 & 236 \\
Shenzen-Validation  & 50 & 50 \\
Shenzen-Test  & 50 & 50 \\ \hline
Montgomery-External Test Set  & 80 & 58 \\
\hline
\end{tabular}
\end{center}
\end{table}

Using the ChexNet and MetaChexnet networks which were trained in Phase I,  we performed fine-tuning on the Shzenzen TB dataset. As a baseline, we consider a DenseNet121 ImageNet pre-trained network and train it on the Shezen dataset. Montgomery test set was used exclusively to study the generalization ability of the networks and was not included in the training scheme. In Table \ref{auc_table_tb} we display our results on the Shenzen test subset and on the Montogemery external set.
We note that the highest AUC on the Shenzen dataset was attained by MetaChexNet.

{\it Generalization Ability }
Previous work \cite{c11} combined the available datasets into a single dataset from which the test subset was extracted. In our experiments, we examine the generalization ability on a dataset containing different population which was acquired using different radiography technology \cite{c7}.
Examining the generalization ability over Montgomery dataset, we notice the high AUC is maintained in ChexNet and MetaChexnet while a decline is evident in ImageNet pre-trained DenseNet121 network. This demonstrates the robustness of the features learned from ChestXray14. In addition, we examine the AUC results on a combined set comprised of Montgomery set and Shenzen test subset (Table \ref{auc_table_tb}). On the combined test-set, ChexNet and MetaChexnet maintained high AUC while a decrease was noted in the ImageNet pre-trained DenseNet121.

\begin{figure}
	\centering
	\fbox{\includegraphics[width=3.2in,trim={0.2cm 0.4cm 1cm 0.2cm}]{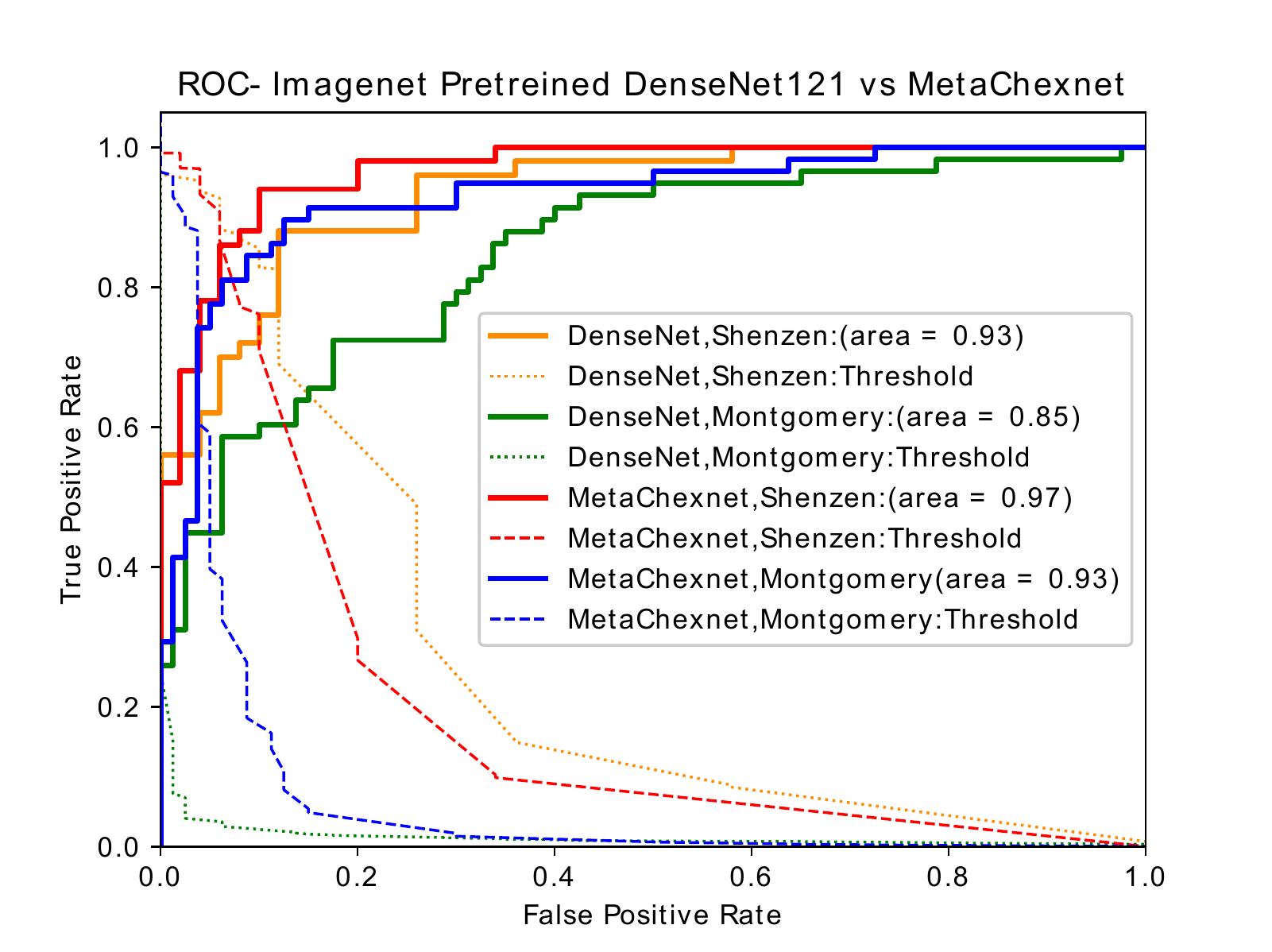}}
	\caption{TB detection ROC graphs for ImageNet pre-trained DenseNet121 \cite{c5} and MetaChexNet (proposed)}
	\label{auc_image_net}
\end{figure}

We attribute that to the different range of output probabilities between the test sets. This lack of uniformity in output probability range inhibits the selection of an optimal threshold that is suitable for the combined set. In Fig.\ref{auc_image_net} the ROC curves of MetaChexNet and the baseline approach are displayed. It can be observed that MetaChexNet demonstrated better performance over the entire range of thresholds.

\begin{table}[ht]

\caption{AUC results for TB classification}
\vspace{-3 ex}
\label{auc_table_tb}
\begin{center}
\begin{tabular}{l|c|c|c|c}
\hline

  & Size & DenseNet121& ChexNet& MetaChexnet \\
  &  & ImageNet pre-trained  &  &  (Proposed) \\
\hline
Shenzen  &100  &0.933  & 0.928 & \textbf{0.965}\\
Montgomery &138 & 0.846 & \textbf{0.952} & 0.928  \\\hline
Combined & 238 &0.803 & \textbf{0.944} & 0.937 \\
\hline
\end{tabular}
\end{center}
\end{table}

\section{Discussion}

We have displayed a scheme for learning image features specific to Chest-Xray by using a hospital-scale dataset with pathology labels and metadata. While pathology labels in the ChestXray14 can be noisy, metadata is an objective and accurate label thus it can be advantageous to the feature learning process. In addition, it contributes information for both normal cases and pathological cases. 
In our experiments on small-scale datasets, we have demonstrated the advantage of our proposed architecture (MetaChexNet) over ImageNet pre-trained architectures in both detection results and generalization ability on an external dataset. 
As an exciting byproduct, we have also demonstrated how a CNN, in addition to pathologies, can predict the patient's metadata using a Chest X-ray as input. One application to that can be the detection of electronic medical record mistakes which can be critical to patient care.


\addtolength{\textheight}{-12cm}   







\begin{thebibliography}{99}

\bibitem{c4}
Rajpurkar, P. et al. ChexNet: Radiologist-level pneumonia detection on chest x-rays with deep learning. arXiv preprint arXiv:1711.05225 (2017).
\bibitem{c6}
Wang, X. et al. ChestC-ray8: Hospital-scale chest x-ray database and benchmarks on weakly-supervised classification and localization of common thorax diseases. In 2017 IEEE Conference on Computer Vision and Pattern Recognition (CVPR).pp. 3462-3471. IEEE. (2017).
\bibitem{c13}
World Health Organization. (‎2018)‎. Global tuberculosis report 2018. World Health Organization. http://www.who.int/iris/handle/10665/274453. 
\bibitem{c11}
Lakhani, P., Baskaran S. "Deep learning at chest radiography: automated classification of pulmonary tuberculosis by using convolutional neural networks." Radiology 284.2 (2017): 574-582.

\bibitem{c10}
Deng, Jia, et al. "ImageNet: A large-scale hierarchical image database." Computer Vision and Pattern Recognition, 2009. CVPR 2009. IEEE Conference on. Ieee, 2009.‏

\bibitem{c12}
Sivaramakrishnan, R., et al. "Comparing deep learning models for population screening using chest radiography." Medical Imaging 2018: Computer-Aided Diagnosis. Vol. 10575. International Society for Optics and Photonics, 2018.‏

\bibitem{c14}
Russakovsky, O., et al. ImageNet Large Scale Visual Recognition Challenge. IJCV, 2015
\bibitem{c5}
Huang, G. et al. Densely connected convolutional networks. arXiv preprint arXiv:1608.06993,
2016

\bibitem{c9}
van der Maaten, L.J.P.; Hinton, G.E. Visualizing High-Dimensional Data
Using t-SNE. Journal of Machine Learning Research 9:2579-2605, 2008.

\bibitem{c8}
Xue, Zhiyun et al. Localizing tuberculosis in chest radiographs with deep learning. 28. 10.1117/12.2293022,2018. 
\bibitem{c7}
Jaeger, S. et al. Two public chest X-ray datasets for computer-aided screening of pulmonary diseases. Quantitative imaging in medicine and surgery, 4(6), 475-7.(2014)


\end{thebibliography}
\end{document}